\documentclass[prb,aps,twocolumn,showpacs,amsmath,amssymb,superscriptaddress,affilletter]{revtex4-1}

\usepackage[usenames,dvipsnames]{color}
\usepackage{graphicx}
\usepackage{tabularx}
\usepackage{slashbox}
\usepackage{multirow}
\usepackage[T1]{fontenc} 
\usepackage[cp1250]{inputenc} 

\newcommand{\be}{\begin{equation}}
\newcommand{\ee}{\end{equation}}

\begin{document}
\title{Dynamics of Charge Leakage From Self-assembled CdTe Quantum Dots}

\author{Ł.~Kłopotowski} \affiliation{Institute of Physics, Polish Academy of Sciences,
al. Lotników 32/46 02-668 Warsaw, Poland}

\author{M.~Goryca} \affiliation{Institute of Experimental Physics, University of Warsaw, ul. Hoża 69, 00-681 Warsaw, Poland}

\author{T.~Kazimierczuk} \affiliation{Institute of Experimental Physics, University of Warsaw, ul. Hoża 69, 00-681 Warsaw, Poland}

\author{P. Kossacki} \affiliation{Institute of Experimental Physics, University of Warsaw, ul. Hoża 69, 00-681 Warsaw, Poland}

\author{P.~Wojnar} \affiliation{Institute of Physics, Polish Academy of Sciences,
al. Lotników 32/46 02-668 Warsaw, Poland}

\author{G.~Karczewski} \affiliation{Institute of Physics, Polish Academy of Sciences,
al. Lotników 32/46 02-668 Warsaw, Poland}

\author{T.~Wojtowicz} \affiliation{Institute of Physics, Polish Academy of Sciences,
al. Lotników 32/46 02-668 Warsaw, Poland}

\date{\today}
\begin{abstract}
We study the leakage dynamics of charge stored in an ensemble of CdTe quantum dots embedded in a field-effect structure. Optically excited electrons are stored and read out by a proper time sequence of bias pulses. We monitor the dynamics of electron loss and find that the rate of the leakage is strongly dependent on time, which we attribute to an optically generated electric field related to the stored charge. A rate equation model quantitatively reproduces the results.
\end{abstract}
\pacs{78.67.Hc, 72.20.Jv, 71.55.Gs}%

\maketitle

Quantum information processing using semiconductor self-assembled quantum dots (QDs) requires a precise control over writting, storage, and readout of electrons from these nanostructures. An experimental testbed for these processes involve QDs embedded in field-effect structures. In such devices, storage of charge carriers was demonstrated and shown to persist over timescales exceeding seconds \cite{lun99} or even hours \cite{fin98}. Optical orientation of carrier spins allowed to realize a programmable QD memory device and exploit it in measurements of longitudinal spin relaxation rates of electrons \cite{kro04} and holes \cite{hei07}. Moreover, storage and readout of charge and spin from a single QD was demonstrated \cite{hei08,hei09}. It has been pointed out that the leakage of charge from these structures stems from thermal \cite{lun99} or tunnel escape \cite{hei09}, photoinduced discharging \cite{hei01}, or capture by deep levels in the barrier \cite{lun99}. These processes result inevitably in an information loss and therefore their role has to be evaluated and carefully taken into account in designing of any future devices.

In this report, we study the escape dynamics of electrons stored in a layer of self-assembled CdTe QDs. We identify the escape mechanism as electron tunneling and propose a theoretical model, which quantitatively reproduces our experimental results. In particular, we address the importance of an optically generated electric field related to the charge stored in the QDs, which substantially modifies the field applied externally and as a result strongly influences the carrier escape. A realization of a QD memory device with II-VI nanostructures is important in view of the possibility of doping these QDs with magnetic ions \cite{mai06}. Indeed, electrical manipulation of the quantum state of a single magnetic ion or a ferromagnetically coupled system of ions would provide new schemes for quantum computation and quantum information storage \cite{fer07,leg09,gor09}.

The samples were grown by molecular beam epitaxy on a (001)-oriented GaAs substrate. A 4 $\mu$m thick p-doped ZnTe buffer layer acted as a back contact. Using a tellurium desorption procedure \cite{tin03}, a single layer of QDs was grown, separated from the back contact by a 15 nm thick Zn$_{0.9}$Mg$_{0.1}$Te spacer layer. QDs were covered with a 100 nm Zn$_{0.9}$Mg$_{0.1}$Te barrier and another Zn$_{0.7}$Mg$_{0.3}$Te blocking barrier to prevent the escape of carriers to the surface. On top, a 10 nm thick, 20 $\mu$m x 100 $\mu$m, semitransparent layer of Ti/Au was evaporated to form a Schottky contact. We note that this structure design is analogous to many InAs/GaAs charge-tunable devices \cite{war00,smi03}.

\begin{figure}[!h]
  \includegraphics[angle=-90,width=.5\textwidth]{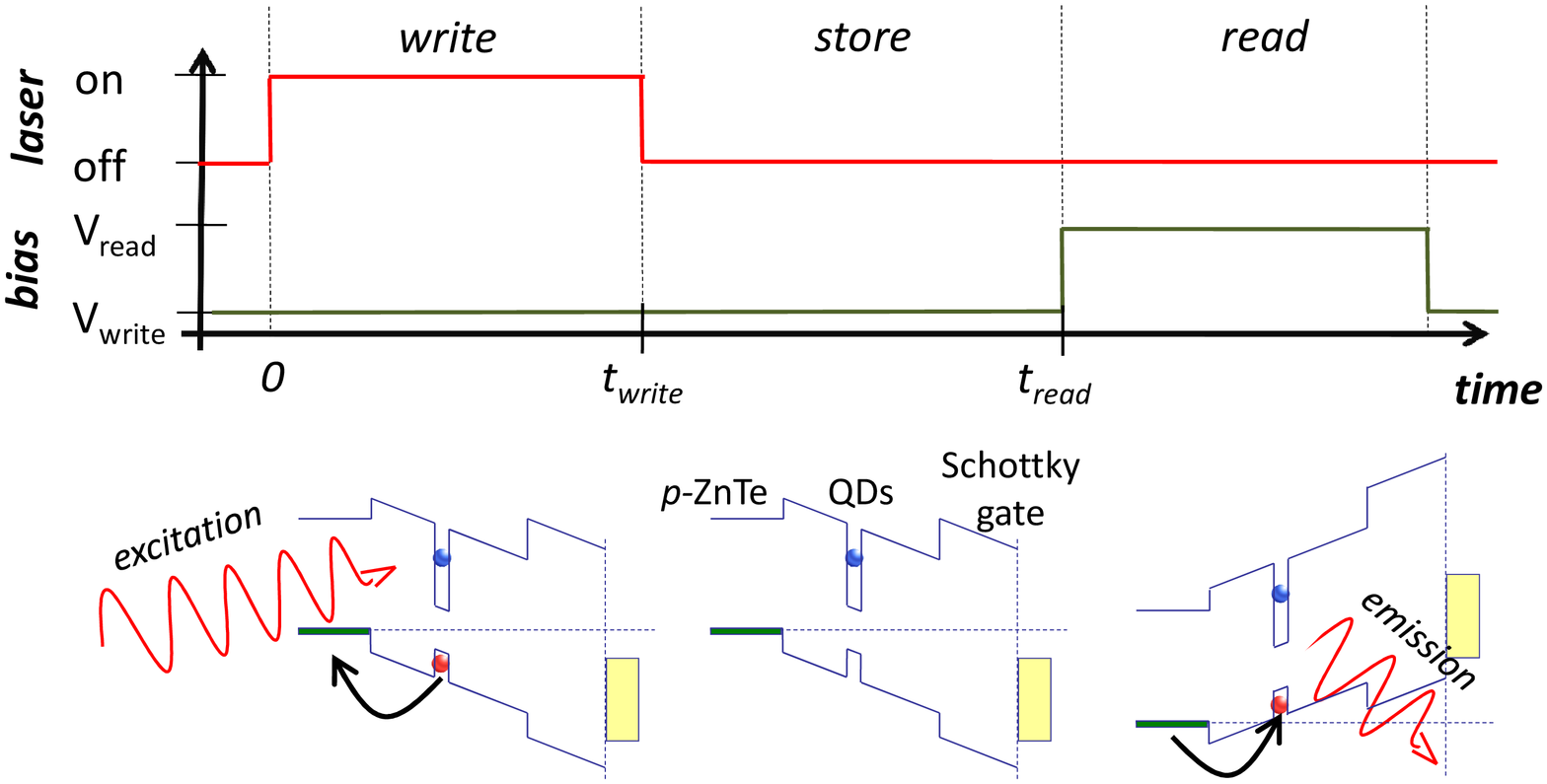}
  \caption{Operation principles of our charge storage device. Top: Schematic of the laser and bias pulse sequence. Bottom: Corresponding band profiles.}
  \label{operation}
\end{figure}

The operation principle of our device is presented in Fig. \ref{operation}. An optical excitation with a laser pulse of a controlled duration, $t_{write}$ in range between 20 ns and 1 ms, and repetition period between 200 ns and 10 ms, creates electron-hole pairs (excitons). The pulses are generated by modulating a cw laser beam with an acousto-optical modulator (AOM). The modulator is synchronized with a pulse generator used to apply the bias voltage. In this {\em writing} phase of the operation cycle, the device is under reverse bias $V_{write}$, which creates an electric field that decreases the triangular barrier between the dots and the back contact, facilitating tunneling of holes out of the dots. As a result, the photons are converted into charge related to electron density in the QD layer. After the laser is switched off, the sample remains under reverse bias with the charge {\em stored} in the QDs. After a controlled delay $\Delta t=t_{read}-t_{write}$ in range between 100 ns and $\sim$ 9 ms, a forward {\em readout} bias $V_{read}$ is applied, resulting in an injection of holes into the dots from the back contact. The resulting recombination is manifested by a distinct readout electroluminescence (EL) peak at $t_{read}$. The duration of the readout pulse is in range between 50 ns and 5 ms. Temporal resolution of our write/store/read setup is about 10 ns limited simultaneously by the fall time of the AOM, the jitter of the generator, and by the characteristic time of the sample electrical response -- determined by its resistance and capacitance and monitored by electrical current measurement with a 500 MHz oscilloscope. The duration of the readout EL is at least ten times shorter than the duration of the readout bias pulse, so we conclude that after readout the dots are emptied of any stored charge. The readout pulse therefore resets the device before the arrival of the next writing laser pulse. Light emission is detected by avalanche photodiodes coupled to a 0.3 m monochromator. The width of the output slit is kept below 1 mm in order to investigate a small subensemble of dots. By tuning the detection energy, we access dots of different sizes, shapes and/or compositions. The signal is time-resolved by a multichannel temporal analyzer providing temporal resolution down to 250 ps. We use two excitation wavelengths: 532 and 472 nm, which correspond to excitation of the QDs below and above the Zn$_{0.9}$Mg$_{0.1}$Te barrier, respectively. The measurements are performed in temperature range between 12 and 20 K.

\begin{figure}[!h]
  \includegraphics[angle=-90,width=.45\textwidth]{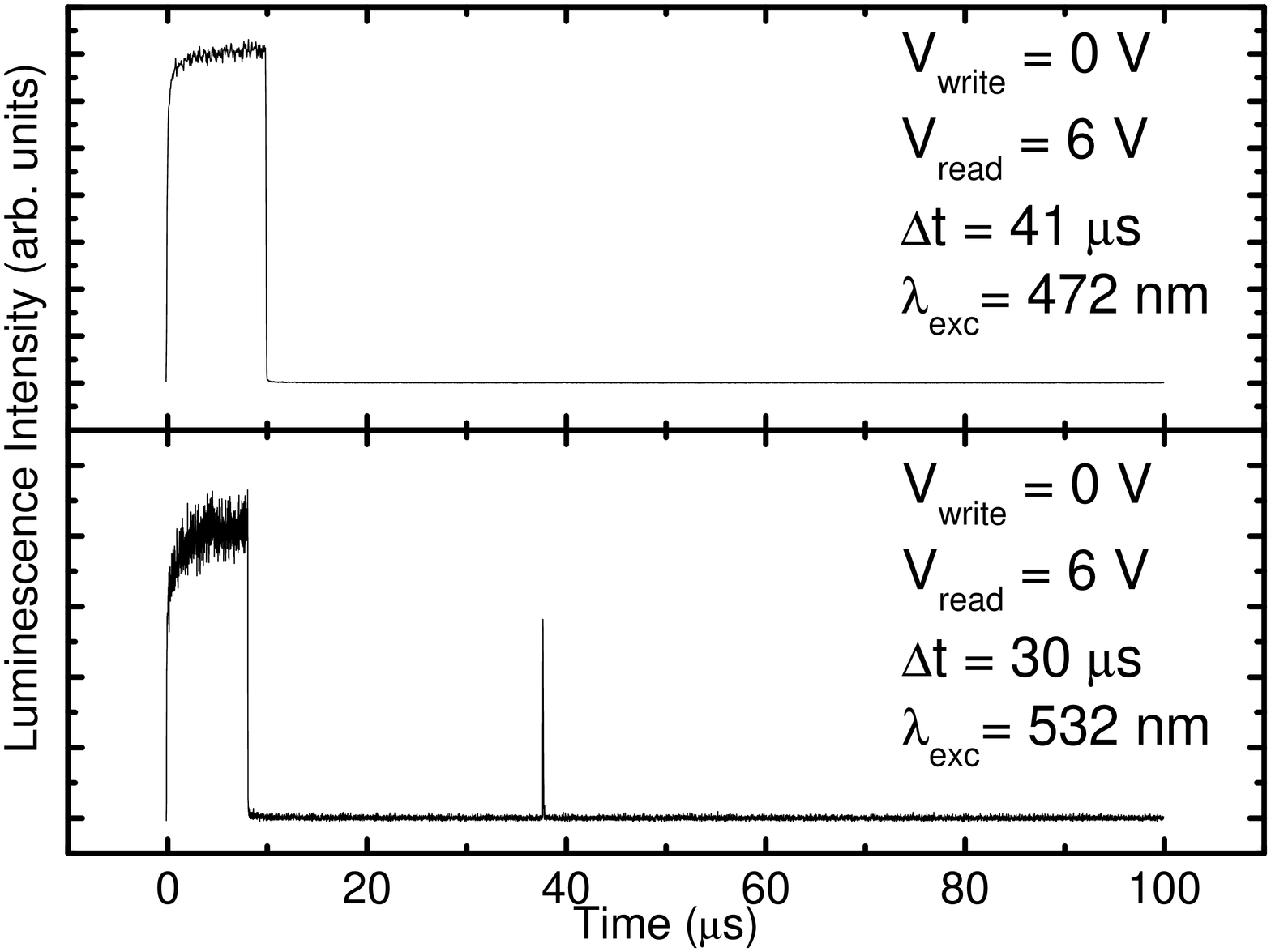}
  \caption{Temporal QD luminescence traces for above (top) and below (bottom) barrier excitation for the write and read biases and read delay given in the annotations.}
  \label{bluegreen}
\end{figure}

Evidence for storage of electrons in the QDs is presented in Fig. \ref{bluegreen}. The top and bottom traces correspond to the above and below barrier excitations, respectively. In both traces we observe a photoluminescence (PL) signal during the illumination of the sample. Since the size of the laser spot on the sample (100 $\mu$m in diameter) is larger than the area of the Schottky contact, this signal results from the excitation of the uncovered part of the sample. In the case of the below barrier excitation, there is also some PL resulting from the competition between hole tunneling and recombination. In that case however, apart from the PL, we detect a clear EL peak exactly at $t_{read}$ corresponding to the application of the read bias. In the case of the above barrier excitation, no EL peak is observed, virtually independent of $V_{write}$ and $V_{read}$\footnote{If $V_{write}$ is close to the flat-band condition, there is some QD capture of electrons from the barrier albeit the resulting readout peaks ar at least an order of magnitude weaker than for a direct QD excitation.}. Indeed, when excited below barrier, the excitons are created directly in the dots. The reverse bias separates electrons from holes, resulting in a photon-to-charge conversion. On the other hand, excitons created with above barrier excitation are dissociated in the Zn$_{0.9}$Mg$_{0.1}$Te barrier and the electrons are dragged by the electric field towards the Zn$_{0.7}$Mg$_{0.3}$Te blocking barrier interface. Since we do not observe any readout peak under the conditions of above barrier excitation, we conclude that no charge is stored neither in the QD layer nor at the barrier/blocking barrier interface \cite{smi03}.

\begin{figure}[!h]
  \includegraphics[angle=-90,width=.5\textwidth]{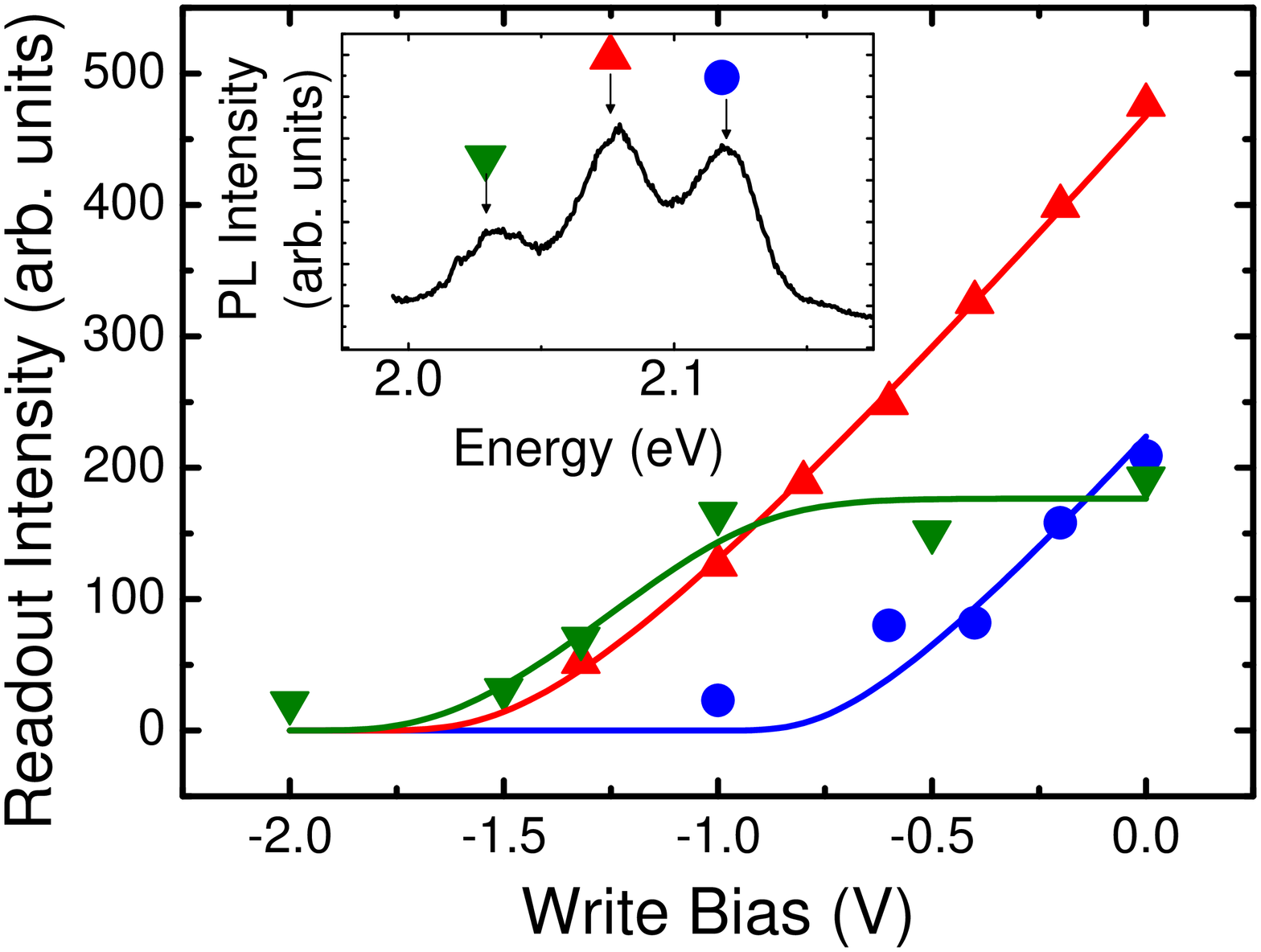}
  \caption{Readout intensity as a function of the reverse write bias for three detection energies. $V_{read}=6$ V, $t_{write}=5$ $\mu$s, $\Delta t$= 10 $\mu$s. Lines are results of model calculations -- see text. Inset: Ensemble PL spectrum with detection energies marked by arrows. The peak structure results from interferences in the buffer layer.}
  \label{bias}
\end{figure}

The traces in Fig. \ref{bluegreen} were collected for $V_{write}=0 $ and $V_{read}=6$ V, respectively\footnote{Under the absence of external voltage, the structure is still under a built-in reverse bias of 1.1 V}. Applying a stronger reverse write bias results in an acceleration of electron leakage from the dots. We demonstrate this behavior in Fig. \ref{bias}. Increasing the reverse bias clearly suppresses the intensity of the readout peak. The dynamics of this decrease is dependent on the QD confinement of the studied subensemble. We find that the readout peak is first suppressed for dots emitting in the high energy part of the ensemble PL (see circles in Fig. \ref{bias}), where we expect dots with a shallow confinement potential. This interpretation is confirmed by the results of a theoretical model (lines in Fig. \ref{bias}) -- see below.

\begin{figure}[!h]
  \includegraphics[angle=-90,width=.45\textwidth]{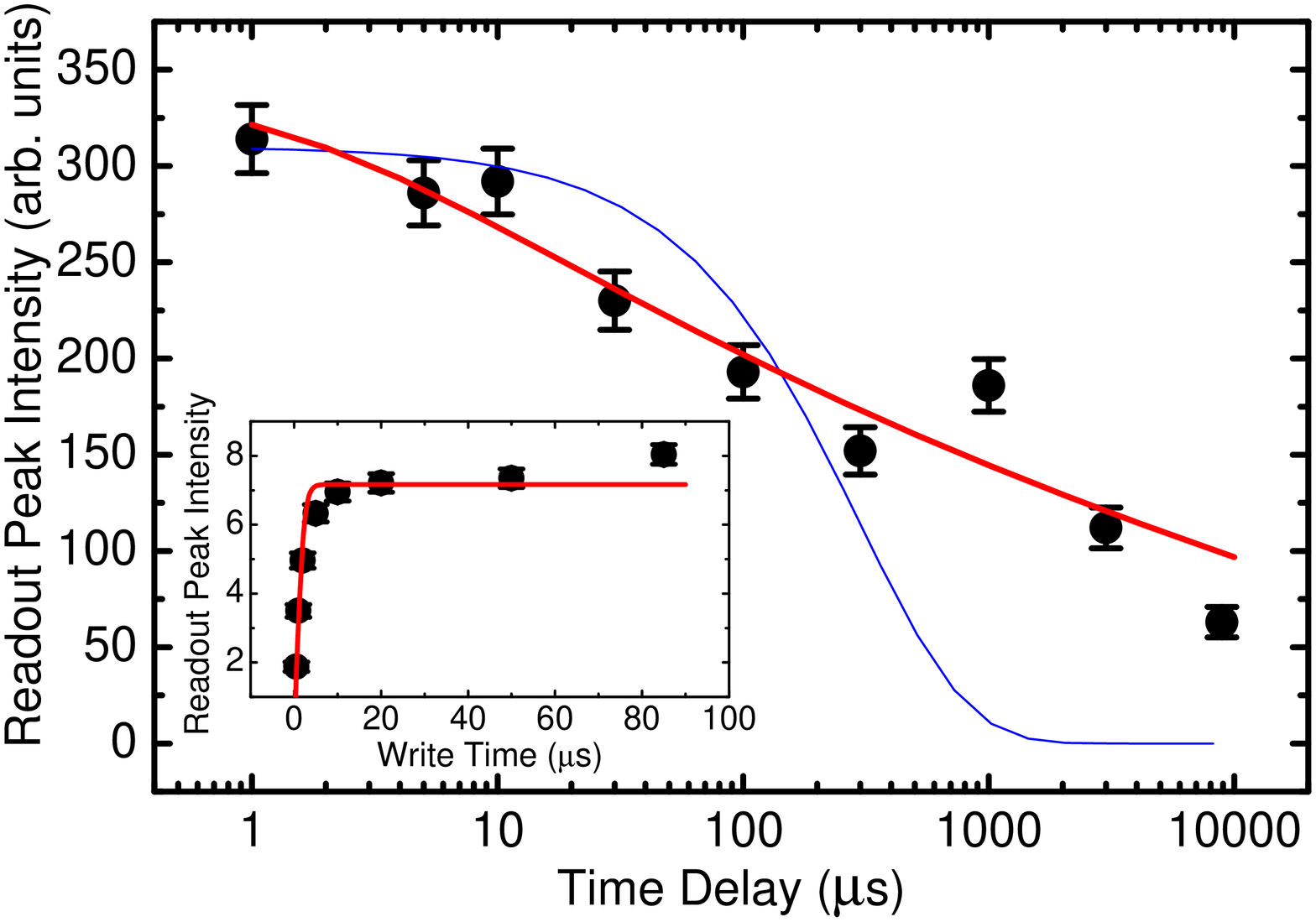}
  \caption{Readout intensity as a function of the time delay between the end of the writing phase of the experiment and the application of the read pulse. $t_{write}=$ 1 ms. Thin blue line represent a monoexponential decay with a characteristic time of 300 ns. Thick red line is a result of model calculations -- see text. Inset: Dependence of readout intensity on the length of the write pulse with $\Delta t=$ 10 $\mu$s. In both cases $V_{write}=0$ V and $V_{read}=3$ V.}
  \label{delay}
\end{figure}

The readout peak intensity decreases also with increasing $\Delta t$. In Fig. \ref{delay}, we present this temporal decay, which directly reflects the leakage of the stored charge. It is clear that the charge persists in the dots for as long as 10 ms. The data in Fig. \ref{delay} was collected for the middle energy subensemble (up-triangles in Fig. \ref{bias}). We suppose that the charge stored in a lower energy subensemble (e.g. down-triangles in Fig. \ref{bias}) persists for even longer. The thin line corresponds to a monoexponential decay and it is clear that the escape process we deal with cannot be described by such a model. Monoexponential decay implies a loss rate constant in time. We find that for short delays the leakage is faster than the exponent in Fig. \ref{delay} and substantially slower for long delays. We then conclude that the leakage mechanisms has a time-dependent efficiency, which has to be self-consistently related to the density of the stored electrons. We propose that the leakage mechanism is electron tunneling out of the dots. Shortly after the excitation pulse, a large density of stored electrons generates an electric field, which decreases the tunnel barrier. As electrons leak out, this electric field disappears and the barrier is restored resulting in slowing down of the escape process.

We check the proposed hypothesis by calculating the tunneling time $t_t$ through a triangular barrier in a semiclassical approach based on a Wentzel-Kramers-Brillouin approximation: $t_t(L, F, E_I)=4 m_e L^2/ \hbar \cdot \exp (4 \sqrt{2 me E_I^3}/3 e F \hbar)$, where $L$, $F$, and $E_I$ are the dot height, electric field, and ionization energy\cite{hel98}. The magnitude of the electric field $F$ depends not only on the applied bias $U$ but also on the number of stored electrons $n(t)$: $F(U,n(t)) = (U-U_b)/w + en(t)/ 2 \varepsilon \varepsilon_0$, where $U_b$ is the built-in bias, $w$ is the width of the space charge layer and $\varepsilon$ is the CdTe static dielectric constant. We assume that the readout peak reflects directly the number of stored electrons, which satisfies the following time-dependent rate equation:

\begin{equation}
\frac{dn(t)}{dt}=G(\alpha,t_{write},t)-\frac{n(t)}{t_r(t_{read},t)}-\frac{n(t)}{t_t(L,n(t),E_I)}
\label{model}
\nonumber
\end{equation}
where $G(\alpha,t_{write},t)$ is a generation function with an efficiency amplitude $\alpha$ and $t_r$ is the recombination time. We assume $t_r=\infty$ in the storage phase, i.e. for $t\leq t_{read}$ and $t_r=300$ ps after the application of the read bias i.e. for $t> t_{read}$\cite{suf06}. We fit the solutions of the above equation to the measured bias dependent readout intensity (Fig. \ref{bias}), to the time delay dependence data (Fig. \ref{delay}) and to check the consistency of our model, to the dependence on $t_{write}$ (Fig. \ref{delay} inset). There are four fitting parameters: $L$, $E_I$, $\alpha$, and a multiplicative scale $\eta$ reflecting the collection efficiency of our setup. In the case of the bias dependent data (Fig. \ref{bias}), we perform the fitting {\em simultaneously} for all three detection energies, keeping $\eta$ and $L$ detection independent. In all cases, we find excellent agreement between the model and the experimental data. Values of the fitted parameters are summarized in Table \ref{param}. Noteworthy, $E_I$ decreases with increasing QD emission energy, as expected. The fitted QD height remains in agreement with the values obtained from TEM studies \cite{klo10}.

\begin{table}[!h]
\begin{tabular}{||c|ccc|c|c||}
  \hline
	&  \multicolumn{3}{c|}{Fig. \ref{bias}}  & Fig. \ref{delay} & Fig. \ref{delay} inset\\ \cline{2-4}
	& \textcolor{ForestGreen}{$\blacktriangledown$} & \textcolor{Red}{$\blacktriangle$} & \textcolor{Blue}{\large{$\bullet$}} & &  \\
  \hline
  $\eta \times 10^{10}$ & \multicolumn{3}{c|}{7.2} & 8.0 & 79 \\
 	$L$ (nm) & \multicolumn{3}{c|}{4.5} & 4.5 & 4.5\\
  $E_I$ (meV) & 225 & 217 & 173 & 180 & 139 \\
  $\alpha \times 10^{-7}$ & 4.90 & 5.84 & 2.67 & 2.9 & 4.2 \\
  \hline
  \end{tabular}
  \caption{Parametrs resulting from fitting of the data in Fig. \ref{bias}, Fig. \ref{delay}, and Fig. \ref{delay} inset. \textcolor{ForestGreen}{$\blacktriangledown$}, \textcolor{Red}{$\blacktriangle$}, and \textcolor{Blue}{\large{$\bullet$}} denote three detection energies from Fig. \ref{bias}. Data from Fig. \ref{bias} is fitted simultaneously, with common $\eta$ and $L$.}
  \label{param}
\end{table}

In summary, we have identified the tunneling as the major mechanism of electron leakage from our CdTe QD memory device. We have shown that the stored charge generates a time-dependent electric field, which decreases the tunnel barrier and influences the leakage rate. At the write bias used in the experiment, this optically generated electric field is about one third of the one applied externally. By calculating the tunneling time self-consistently, taking into account the electron leakage, we quantitatively reproduce the measured decay as a function of time and bias

  The authors wish to thank drs. A. Kudelski and O. Krebs for their help with sample processing.

  This work was partially supported by Polish Ministry of Science and Education grant no. 0634/BH03/2007/33 and by the European Union within the European Regional Development Fund.

%

\end{document}